\documentclass[11pt]{article}
\usepackage[textwidth=15.2cm,textheight=22cm]{geometry}
\usepackage{amsmath,amssymb}
\usepackage{latexsym}
\usepackage{multicol}
\usepackage{graphicx}
\usepackage{bm}
\allowdisplaybreaks[1]

\newcommand{\del}{\partial}
\newcommand{\be}{\begin{equation}}
\newcommand{\ee}{\end{equation}}
\newcommand{\ba}{\begin{eqnarray}}
\newcommand{\ea}{\end{eqnarray}}
\newcommand{\bdm}{\begin{displaymath}}
\newcommand{\edm}{\end{displaymath}}

\def\ba{\bar A}

\def\beq{\begin{equation}}
\def\eeq{\end{equation}}
\newcommand{\half}{\frac{1}{2}}
\newcommand{\nn}{\nonumber}

\newcommand{\ndt}{\noindent}

\def\bea{\begin{eqnarray}}
\def\eea{\end{eqnarray}}
\def\beas{\begin{eqnarray*}}
\def\eeas{\end{eqnarray*}}
\def\sla{\raise.15ex\hbox{$/$}\kern-.57em}

\def\spa#1.#2{\left\langle#1\,#2\right\rangle}
\def\spb#1.#2{\left[#1\,#2\right]}

\begin{document}

\begin{titlepage}
\begin{flushright}    
{\small $\,$}
\end{flushright}
\vskip 1cm
\centerline{\Large{\bf{Fermi-Bose cubic couplings}}}
\vskip 0.5cm
\centerline{\Large{\bf{in light-cone field theories}}}
\vskip 1.5cm
\centerline{Y. S. Akshay and Sudarshan Ananth}
\vskip .5cm
\centerline{\it {Indian Institute of Science Education and Research}}
\centerline{\it {Pune 411008, India}}
\vskip 1.5cm
\centerline{\bf {Abstract}}
We derive light-cone cubic interaction vertices involving fermions and bosons of arbitrary spin by demanding closure of the Poincar\'e algebra. We derive the three-point scattering amplitude corresponding to these interaction vertices and find that they possess interesting factorization properties identical to the case of three boson scattering. 
\end{titlepage}

\section{Introduction}

One aim of this paper is to develop a framework that allows us to derive, from first principles, Lagrangians describing interacting fields of arbitrary spin. The motivation for this symmetry-based approach stems from the early work in~\cite{BBB} where the authors derived cubic self-interaction vertices for bosonic fields. We revisited their work~\cite{SA} motivated by (a) recent developments~\cite{bern}\cite{dixon} in the spinor-helicity approach to scattering amplitudes and (b) to extend their approach to more general cubic vertices. This~\cite{AA} yielded a Lagrangian origin for the KLT relations~\cite{KLT} and more general factorization properties. In this paper, we extend this framework to include fermions, scattering amplitudes involving matter fields and thereby find that they too exhibit similar factorization properties. Some of our results overlap, as expected, with those in~\cite{elvang,smat}. \\

A more general aim is to develop a framework where we can, potentially, derive consistent Lagrangians describing interacting fields of higher spin. One motivation for this is to explicitly see the no-go theorems~\cite{BBS}, regarding interacting higher-spin fields, in action. In other words, we should see the breakdown of this symmetry based approach at the quartic level. However, the approach presented here sacrifices both manifest locality and manifest Lorentz invariance in order to focus on the physical helicity states. Since most no-go theorems assume both manifest locality and Lorentz invariance this issue may prove interesting. \\

Another reason why this framework is worth developing is to study aribtrary spin theories in non-flat space-time backgrounds. This framework may be extended easily to $AdS_4$ (for related results, see~\cite{AAM} for the case of spin$=2$). At the quartic vertex level, this approach could lead us to the key ingredients for the (Fronsdal) Vasiliev program~\cite{MV}. In other words, an attempt to derive a consistent quartic interaction vertex involving higher spin fields in $AdS_4$ should force us to include an infinite tower of higher spin fields thus providing a Lagrangian origin to the Vasiliev equations of motion.\\

An additional utility of this procedure is to attempt a construction of the elusive $\mathcal{N}=(2,0)$ theory in 6 dimensions~\cite{Ramond}. This can be done by closing the 6 dimensional superconformal algebra . Such an approach has proved successful in the past. See~\cite{PSU} for a derivation of $\mathcal{N}=4$ Yang-Mills  by closing the PSU(2,2$|$4) algebra.  

\section{Free arbitrary spin fields in light-cone gauge}
Manifest Lorentz covariance in describing massless fields in (3+1) dimensions requires the use of tensors or spinor-tensors of appropriate rank. Since massless fields in (3+1) dimensions have only two physical degrees of freedom, preserving manifest covariance comes at the cost of introducing unphysical degrees of freedom. Consequently, constructing interactions in such a formalism is cumbersome. A light-cone gauge construction of interaction vertices is more natural as it deals exclusively with the two physical degrees of freedom. However, this description sacrifices manifest covariance so closure of the Poincar\'e algebra must be enforced. \\

\subsection{Free Bosonic fields}
In this section, we briefly outline the free theory of massless bosonic fields of arbitrary spin in light-cone gauge. Light-cone co-ordinates in Minkowski space-time with signature $(-,+,+,+)$ are defined by
\begin{eqnarray}
x^{\pm}=\frac{x^{0}\pm x^{3}}{\sqrt{2}} \;,\qquad
x = \frac{x^{1}+ix^{2}}{\sqrt{2}} \;,\qquad\bar{x}= \frac{x^{1}-ix^{2}}{\sqrt{2}}\,
\end{eqnarray}
with the corresponding derivatives being $\partial_{\pm}\,,\,\,\bar{\partial}$ and $\partial$. \\
The physical degrees of freedom in this theory are the positive and negative helicity states. We will denote these by $\phi$ and $\bar{\phi}$ respectively. The value of the helicity is determined by the action of the rotation generators on them. The Action describing a massless, spin-$\lambda$ bosonic field is 
\bea
S = \int d^{4}x \,\,\,\half\,\bar\phi\,\Box\,\phi 
\eea
The Poincar\'{e} generators acting on free fields are
\bea
\delta_{p^{-}}\phi=i\frac{\partial\bar{\partial}}{\partial_{-}}\phi=-\delta_{p_+}\phi \qquad \delta_{p^+}\phi=-i\partial^{+}\phi=-\delta_{p_-}\phi \qquad  \delta_{p}\phi=-i\partial\phi ,
\eea
\begin{eqnarray}
\delta_{j}\!\!\!\!\!\!&\phi= i(x\bar{\partial}-\bar{x}\partial - \lambda)\phi \;,\qquad  &\delta_{j^{+}}\phi = (x^+ \partial-x\partial^+)\phi , \nn\\
\delta_{j^{+-}}\phi\!\!\!\!&=(x^{+}\frac{\partial\bar{\partial}}{\partial^+}-x^-\partial^+ )\phi \;,\qquad & \delta_{j^-}\phi=(x^{-}\partial-x\frac{\partial\bar{\partial}}{\partial^+}+\lambda \frac{\partial}{\partial^+})\phi \
\end{eqnarray} 
The three remaining generators are the appropriate complex conjugates. We follow the pole prescription in~\cite{SM} to define the non local operator $\frac{1}{\del^+}$.

\subsection{Free fermionic fields}
We can similarly describe free fermions in light-cone gauge by building suitable projection operators to eliminate the unphysical degrees of freedom. We will work with a sinlge component Grassmann valued field. Details regarding this are presented in Appendix A. The light-cone Action for fermionic fields of half-integer spin $\lambda$ is
\bea
\label{fermion}
S = \int d^{4}x\,\,\, i \bar\psi\,\frac{\Box}{\del_-}\,\psi .
\eea
The corresponding Poincar\'e generators represented on fermionic fields are
\bea
\delta_{j}\!\!\!\!\!\!&\psi= i(x\bar{\partial}-\bar{x}\partial + \lambda)\psi \;,\qquad  &\delta_{j^{+}}\psi = (x^+ \partial-x\partial^+)\psi , \nn\\
\delta_{j^{+-}}\,\,\,\!\!\!\!\psi &=(x^{+}\frac{\partial\bar{\partial}}{\partial^+}-x^-\partial^+ -\half)\psi \;,\qquad & \delta_{j^-}\psi=(x^{-}\partial-x\frac{\partial\bar{\partial}}{\partial^+}+(\lambda +\half) \frac{\partial}{\partial^+})\psi \
\eea
and their complex conjugates. Note that $\bar{\psi}$ carries helicity of $\lambda$ which is a positive half integer, in contrast with the bosonic case where $\bar{\phi}$ carried negative helicity. Another difference, from the bosonic case, is the presence of a third piece in $\delta_{j^{+-}}$.\\
\section{Interacting arbitrary spin fields} 
To begin the construction of interaction vertices, we start with the Hamiltonian for a free theory involving a massless boson of spin $\lambda$ and a massless fermion of spin $\lambda'$.
\begin{equation}
\label{hamil}
H = \int d^3x\,\,\,\left(-\bar\phi\,\partial\bar\partial\,\phi\ +i \bar\psi\,\frac{\partial\bar\partial}{\del_-}\,\psi\right)=\int d^{3}x\,\,\left(\partial_{-}\bar{\phi}\,\delta_{\mathcal{H}}\phi + i \bar\psi\,\delta_{\mathcal{H}}\,\psi\right)
\end{equation}
where
\begin{eqnarray}
\delta_{\mathcal{H}}\phi\equiv\lbrace \phi,\mathcal{H}\rbrace=\partial_{+}\phi .
\end{eqnarray}
Here $\lbrace ,\rbrace$ denotes the Poisson bracket.\\\\
\ndt
To describe an interacting theory we add non-linear terms order by order in a coupling constant $\alpha$ to the Hamiltonian. From the Poincar\'e algebra, it is clear that some of the generators also pick up non linear corrections. These are the dynamical generators (involving $\del^-$) and the form of their corrections is
\begin{eqnarray}
\label{corr}
\delta_{j^{+-}}\phi \!\!\!\!\!\!\!\!\!&&= \delta_{j^{+-}}^{0}\phi -ix^{+}\delta^{\alpha}_{\mathcal{H}}\phi + O(\alpha^{2})\ ,\nn \\
\delta_{j^{-}}\phi \!\!\!\!\!\!\!\!\!&&= \delta_{j^{-}}^{0}\phi +ix\delta^{\alpha}_{\mathcal{H}}\phi +\delta^{\alpha}_{s}\phi + O(\alpha^{2})\ , \nn \\
\delta_{\bar{j}^{-}}\phi \!\!\!\!\!\!\!\!\!&&= \delta_{\bar{j}^{-}}^{0}\phi +i\bar{x}\delta^{\alpha}_{\mathcal{H}}\phi +\delta^{\alpha}_{\bar{s}}\phi + O(\alpha^{2}),
\end{eqnarray}
where $\delta_{\mathcal{H}}^\alpha$ is the first order correction to $\delta_{\mathcal{H}}$. $\delta_{s}^{\alpha}$ and $\delta_{\bar{s}}^{\alpha}$ represent corrections to the spin part of the transformations (irrelevant to calculations presented here~\cite{BBB}).\\

\ndt We make a preliminary Ansatz for $\delta_{\mathcal{H}}^\alpha \phi_1$ of the skeletal form  $\bar{\phi}_1\psi_2\psi_3$ where the fields $\phi_1$, $\psi_2$ and $\psi_3$ have spins $\lambda_1$, $\lambda_2$ and $\lambda_3$ respectively. $\lambda_1$ is an integer while $\lambda_2$ and $\lambda_3$ are half-integers.  From (\ref{corr}), we can see the corrections to all other dynamical generators can be expressed in terms of $\delta_{\mathcal{H}}^{\alpha}\phi$. The precise form of the Ansatz is
\begin{equation}
\label{integers}
\delta^{\alpha}_{\mathcal{H}}\phi_1=\alpha\,A\,\partial^{+\mu}\left[\partial^a\partial^{+\sigma}\psi_2\partial^b\partial^{+\rho}\psi_3\right]\  ,
\end{equation}
where $\mu ,\rho ,\sigma ,a,b$ are integers and $A$ is a numerical factor that could depend on the variables and spins.\\\\
\ndt
To ensure that the theory resulting from such a vertex is Lorentz covariant, we demand the closure of the Poincar\'e  algebra at this order in $\alpha$. We begin by demanding closure with the simplest kinematical generators.
\begin{eqnarray}
&&\left[\delta_{j},\delta_{\mathcal{H}}^{\alpha}\right]\phi_1 =0\ \hspace{1cm}\left[\delta_{j^{+-}},\delta_{\mathcal{H}}\right]^{\alpha}\phi_1\nn =-\delta^\alpha_{\mathcal{H}}\phi_1\ ,
\end{eqnarray}
which impose the following constraints on $a, b, \mu, \rho$ and $\sigma$.
\begin{eqnarray}
\label{conds1}
\nn a+b\!\!\!\!\!\!&&=\lambda_1+\lambda_2+\lambda_3\equiv \lambda  \\
\mu +\rho +\sigma\!\!\!\!\!\!&&=-2 \ .
\end{eqnarray} 

\ndt We then demand, 
\bea
\left[\delta_j^{-}, \delta_{H}\right]^{\alpha}\phi_{1} =0 \qquad \left[\delta_{\bar j^{+}}, \delta_{H}\right]^{\alpha}\phi_{1} = 0\: ,
\eea
which yield the following constraints,  
\bea
\label{conds3}
\nn \sum_{n=0}^{\lambda}\!\!\!\!\!\!&& A_n\,{\biggl \{}\;-(\mu_n +1+\lambda_1)\del^{+(\mu_n -1)}\del(\del^n\del^{+\rho_n}\psi_2\del^{(\lambda -n)}\del^{+\sigma_n}\psi_3)\\
&&-(\sigma_n +\lambda_2+\half)\del^{+\mu_n} (\del^{(n+1)}\del^{+(\sigma_n -1)}\psi_2 \del^{(\lambda -n)}\del^{+\rho_n}\psi_3)\\	
\nn &&-(\rho_n +\lambda_3+\half)\del^{+\mu_n} (\del^n\del^{+\sigma_n}\psi_2 \del^{(\lambda-n+1)}\del^{+(\rho_n -1)}\psi_3){\biggr \}}\ =0\ ,\\
\nn\sum_{n=0}^{\lambda}\!\!\!\!\!\!&& A_n{\biggl \{}\;n\,\del^{+\mu_n}(\del^{(n-1)}\del^{+(\sigma_n +1)}\psi_{2} \del^{(\lambda -n)}\del^{+\rho_n}\psi_{3})\\&&+(\lambda -n)\del^{+\mu_n}(\del^{n}\del^{+\sigma_n}\psi_{2}\bar{\del}^{(\lambda -n-1)}\del^{+(\rho_n +1)}\psi_{3}){\biggr \}}=0\ .
\eea
\\
The solution to these constraints is given below. As these calculations are similar to the bosonic case, we refer the reader to~\cite{AA} for further details. 
\be
\label{A}
\nn A_{n}= (-1)^{(n)} \binom{\lambda}{n};\qquad
\sigma_n = -\lambda - \lambda_2 +n-\half \ ; \qquad \rho_n = n-(\lambda_3+\half) \ ;\qquad \mu_n=-(\lambda_1+1)\ .
\ee
\\
Thus (\ref{integers}) reads
\be
\delta_H^\alpha \phi_1=\alpha\sum_{n=0}^{\lambda}(-1)^n \binom{\lambda}{n}\,\frac{1}{\partial^{+(\lambda_1 +1)}}\left[\partial^n\frac{\partial^{+(\lambda-n)}}{\partial^{+(\lambda_2+\half)}}\psi_2\,\partial^{(\lambda -n)}\frac{\partial^{+n}}{\partial^{+(\lambda_3+\half)}}\psi_3\right]\: .
\ee
Using this, from (\ref{hamil}) we obtain the Hamiltonian and thus the Action 
\bea
\label{Action}
S = \int d^4x\hspace{-.5cm} &&\left[\half\bar\phi_1\Box\phi_2 + i \bar\psi_2\,\frac{\Box}{\del_-}\,\psi_2 + i \bar\psi_3\,\frac{\Box}{\del_-}\,\psi_3\right.\\
\nn 
&&\left. + \,\alpha \bar{\phi}_1\sum_{n=0}^{\lambda}(-1)^n \binom{\lambda}{n}\,\frac{1}{\partial^{+\lambda_1}}\left[\partial^n\frac{\partial^{+(\lambda-n)}}{\partial^{+(\lambda_2+\half)}}\psi_2\,\partial^{(\lambda -n)}\frac{\partial^{+n}}{\partial^{+(\lambda_3+\half)}}\psi_3\right]\right]
\eea
This Action may be supplemented with a variety of other interactions. In particular, the algebra tells us that if $\phi_1$ has odd helicity and is to have cubic self-interactions, then it must carry an internal symmetry group with an antisymmetric structure constant~\cite{AA}. We can thus have two kinds of vertices  having the schematic forms $t_a\bar{\phi}_1^a\psi_2^b\psi_{3b}$ and $\bar{\phi}_1^a\psi_2^b\psi_3^cf_{abc}$. The first of these has the bosonic field in the Adjoint representation of the gauge group($t_a$ is its generator) with the fermionic fields in any other representation. The second vertex has all three fields in the same representation and linked by an antisymmetric structure constant.\\\\
\ndt
For example, consider the case where $(\lambda_1,\lambda_2,\lambda_3)=(1, \half, -\half)$. With an SU(3) internal symmetry, the first kind of vertex is the coupling of gluons to quarks in QCD with the second variety corresponding to the cubic coupling in $\mathcal{N}=4$ SYM. If we compare this second type of vertex, we find an exact match with ~\cite{SYM} (line 5, equation 3.13). 
   
\section{Relation to scattering amplitudes}
Wel now evaluate the three point amplitude resulting from the vertex in (\ref{Action}). We introduce spinor helicity notation following the conventions in~\cite{dixon}
\begin{equation}
<kl>= \sqrt{2}\:\frac{(kl_{-}-lk_{-})}{\sqrt{k_{-}l_{-}}} \qquad  [kl]= \sqrt{2}\:\frac{(\bar{k}l_{-}-\bar{l}k_{-})}{\sqrt{k_{-}l_{-}}}\  .
\end{equation}
The cubic vertex from (\ref{Action}) in momentum space is
\be
\label{N=4}
\alpha\int \frac{d^4p}{(2\pi)^4}\frac{d^4k}{(2\pi)^4}\frac{d^4l}{(2\pi)^4}(2\pi)^4 \delta^4(p+k+l)\frac{(k_-l-l_-k)^\lambda}{p_-^{\lambda _1}k_-^{\lambda _2+\half}l_-^{\lambda _3+\half}}\tilde{\bar{\phi}}_1(p)\tilde{\psi}_2(k)\tilde{\psi}_3(l)\ ,
\ee
\ndt
To obtain the amplitude, we simply put all three fields on mass shell. Along with the constraint ($p^2=0$) on the momenta, this yields a numerical factor for bosonic fields and a factor proportional to the square root of momentum for fermionic fields (from Appendix A). The final amplitude has the following form
\bea
\label{final}
<pk>^{^{(-\lambda_1 +\lambda_2 -\lambda_3)}}\:\:<kl>^{^{(\lambda_1 +\lambda_2 +\lambda_3)}}\:\:<lp>^{^{(-\lambda_1 -\lambda_2 +\lambda_3)}}.
\eea
This has exactly the same form as the amplitude for scattering of three bosons derived in~\cite{AA}. This is consistent with the general result for three point amplitudes derived in~\cite{smat} using S-matrix arguments, in~\cite{elvang} using little group scaling and ~\cite{RRM, bengt} using a Fock-space approach.

\vskip .4cm
\begin{center}
* ~ * ~ *
\end{center}

\ndt 
In contrast to the purely bosonic case~\cite{AA}, an exact factorization in the fermionic case only happens at the level of the scattering amplitudes. Nevertheless the Lagrangian in (\ref{Action}) provides an off shell origin for this generalized KLT-like property. It would be interesting to move beyond the tree level approximation and discover if such factorizations hold at one and higher loop levels.
\vskip 0.5cm
\ndt {\it {Acknowledgments}}
\vskip 0.3cm

\ndt We thank Hidehiko Shimada for discussions. We thank the anonymous referee of~\cite{AA}, Bernard de Wit and Senthil Todadri for asking us questions about the incorporation of fermions into this framework. YSA is supported by an INSPIRE fellowship from the Department of Science and Technology, Government of India.

\section*{Appendix A}
In this appendix, we present the details of light-cone reduction of spinors. We work in a spacetime with metric signature $(-+++)$. The $4\times 4$ Gamma matrices are
\bea
\gamma^0 = \left(\begin{array}{cc}
0 & 1 \\
1 & 0 \end{array}\right)
\hspace{1cm}\gamma^i = \left(\begin{array}{cc}
0 & \sigma^i \\
-\sigma^i & 0 \end{array}\right)
\hspace{1cm}C = \left(\begin{array}{cc}
i\sigma^2 & 0 \\
0 & -i\sigma^2 \end{array}\right).
\eea
At the heart of the light-cone formalism is the separation of dynamical and kinematical degrees of freedom. To this end, we define the following matrices
\bea
P_+ \equiv \half \gamma_+\gamma_- = \left(\begin{array}{cccc}
0 & 0 & 0 & 0 \\
0 & 1 & 0 & 0 \\
0 & 0 & 1 & 0 \\
0 & 0 & 0 & 0 \\
\end{array}\right)
\qquad
P_- \equiv \half \gamma_-\gamma_+ = \left(\begin{array}{cccc}
1 & 0 & 0 & 0 \\
0 & 0 & 0 & 0 \\
0 & 0 & 0 & 0 \\
0 & 0 & 0 & 1 \\
\end{array}\right)
\eea
where
\bea
\nn
&&\gamma_+ = \frac{1}{\sqrt{2}}\left(\gamma_0+\gamma_3\right) = \frac{1}{\sqrt{2}}\left(\begin{array}{cc}
0 & -1+\sigma^3 \\
-1-\sigma^3 & 0 \end{array}\right)\\
&&\gamma_- = \frac{1}{\sqrt{2}}\left(\gamma_0-\gamma_3\right) = \frac{1}{\sqrt{2}}\left(\begin{array}{cc}
0 & -1-\sigma^3 \\
-1+\sigma^3 & 0 \end{array}\right).
\eea
It is easy to see that they are projection operators as they satisfy
\bea
\nn
P_+^2 = P_+ \qquad P_-^2 = P_- \qquad P_+P_-=0 \qquad P_++P_-=1.
\eea
We can now begin the procedure of light-cone reduction. Let
\be
\Psi = \left(\begin{array}{c}
\chi_1\\
\chi_2\\
\chi_3\\
\chi_4 \end{array}\right)
\ee
be a Dirac spinor with Grassmann valued components. Imposing the Majorana condition $\Psi = C\bar{\Psi}^T$ yields
\bea
\label{maj}
\chi_1 = \bar{\chi}_4 \qquad\ ; \qquad \chi_2 = -\bar{\chi}_3\ .
\eea
Define 
\bea
\Psi_+ = P_+\Psi \hspace{0.5cm};\hspace{0.5cm} \Psi_- = P_-\Psi\ .
\eea
Acting the projection operators on the Dirac equation yields the following two equations
\bea
\label{kin}
\del_-\Psi_- = \half \gamma_-\gamma_i\del_i\Psi_+\ , \qquad\qquad i = 1, 2
\eea
\bea
\label{dyn}
\del_+\Psi_+ = \half \gamma_+\gamma_i\del_i\Psi_-\ , \qquad\qquad i = 1, 2
\eea
This first of these is kinematical and can be solved to yield 
\be
\label{plus}
\chi_1 = \frac{\bar{\del}}{\del_-}\chi_2\ .
\ee
The entire spinor can now be written in terms of $\chi_3 \equiv \chi$ as
\be
\label{reduced}
\Psi = \left(\begin{array}{c}
-\frac{\bar{\del}}{\del_-}\bar{\chi} \\
-\bar{\chi}\\
\chi\\
-\frac{\del}{\del_-}\chi \end{array}\right)\ .
\ee
This completes the reduction procedure. Using (\ref {reduced}) in the Dirac Lagrangian and keeping track of Grassmann signs, we arrive at the form (\ref{fermion}).\\
To calculate scattering amplitudes, we need to solve the dynamical equation (\ref{dyn}). Use of (\ref{kin}) in (\ref{dyn}), yields the Klein-Gordon equation as expected. The positive frequency solution is
\bea
\Psi = u(p)e^{ipx} = \left(\begin{array}{c}
u_1\\
u_2\\
u_3\\
u_4 \end{array}\right)e^{ipx} \  .
\eea  
Imposing the constraints (\ref{maj}) and (\ref{plus}) yields
\bea
\Psi =\left(\begin{array}{c}
-\frac{\bar{p}}{p_-} \bar{u}_3\\
-\bar{u}_3\\
u_3\\
-\frac{p}{p_-}u_3 \end{array}\right)e^{ipx}\ .
\eea  
We normalize the spinor using
\bea
\Psi^{\dagger}_+\Psi_+ = \sqrt{2}p_- \implies 2|u_3|^2 = \sqrt{2}p_-\ ,
\eea
arriving at 
\bea
\Psi = 2^{\frac{-1}{4}}p_-^{-\half}\left(\begin{array}{c}
\bar{p}\\
-p_-\\
p_-\\
-p \end{array}\right)e^{ipx}\ . 
\eea  
This also applies to the negative frequency spinor.

\end{document}